\documentclass[preprint,tightenlines,aps,floats,nofootinbib,amssymb]{revtex4}
\usepackage{epsf,epsfig}

\newcommand{\be}{\begin{equation}}
\newcommand{\ee}{\end{equation}}
\newcommand{\bea}{\begin{eqnarray}}
\newcommand{\eea}{\end{eqnarray}}

\newcommand{\lsim}{
\mathrel{\hbox{\rlap{\hbox{\lower4pt\hbox{$\sim$}}}\hbox{$<$}}}}
\newcommand{\gsim}{
\mathrel{\hbox{\rlap{\hbox{\lower4pt\hbox{$\sim$}}}\hbox{$>$}}}}

\preprint{
\hbox to \hsize{
\hfill$\vcenter{\hbox{\bf MADPH-05-1423}
            \hbox{\bf UPR-1121T}
                \hbox{\bf hep-ph/0508027}
                \hbox{August 2005}}$ }
}

\begin{document}

\title{\vspace*{.75in}
Lightest Neutralino in Extensions of the MSSM}

\author{
Vernon Barger$^1$, Paul Langacker$^2$, and Hye-Sung Lee$^1$}

\affiliation{
$^1$Department of Physics, University of Wisconsin,
Madison, WI 53706 \\
$^2$Department of Physics and Astronomy, University of Pennsylvania,
Philadelphia, PA 19104
\vspace*{.5in}}

\thispagestyle{empty}

\begin{abstract}
\noindent We study neutralino sectors in extensions of the MSSM that
dynamically generate the $\mu$-term. The extra neutralino states are
superpartners of the Higgs singlets and/or additional gauge bosons.
The extended models may have distinct lightest neutralino properties
which can have important influences on their phenomenology. We
consider constraints on the lightest neutralino from LEP, Tevatron,
and $(g-2)_\mu$ measurements and the relic density of the dark
matter. The lightest neutralino can be extremely light and/or
dominated by its singlino component which does not couple directly
to SM particles except Higgs doublets.
\end{abstract}
\maketitle

\newpage

\section{Introduction}
\label{introduction}
One of the most important challenges to the Standard Model (SM) of
particle physics is the cosmological observation that the dominant
component of matter in the universe is unexplained. The preferred
interpretation of the observational data is that this matter is
massive (cold), electrically neutral (dark), and also stable or
long-lived. The Supersymmetric extension of the SM provides a
well-motivated candidate for cold dark matter. Supersymmetry (SUSY)
predicts the superpartners of the neutral gauge bosons and the
neutral Higgs bosons with masses of electroweak (EW) scale and
couplings of EW strength. The lightest neutralino is the lightest
Supersymmetric particle (LSP) over most of the parameter space.
Assuming $R$-parity conservation that prevents rapid proton decay,
the lightest neutralino is stable.

The Minimal version of the Supersymmetric SM (MSSM) assumes
$R$-parity conservation and has an extra Higgs doublet and general
SUSY breaking soft terms \cite{MSSMreview}. Extensive studies of the
MSSM show that its lightest neutralino has the right ranges of mass
and interaction strength to be a good cold dark matter (CDM)
candidate \cite{MSSMCDM}. Although the MSSM may be the optimal low
energy Supersymmetric model with minimal extension of the fields and
symmetry, the model may not fully describe the TeV scale physics. It
is important to see if other versions of the Supersymmetric SM can
also give acceptable CDM.

A good starting point for alternatives to the MSSM is its
theoretical weakness. The MSSM has a fine-tuning problem, the so
called $\mu$-problem \cite{muproblem}. The $\mu$-term is the only
dimensionful parameter in the SUSY conserving sector of the MSSM,
and it is required to have the same scale as the SUSY breaking
parameters (EW/TeV-scale soft terms) to give an EW scale Higgs
vacuum expectation value (VEV) without fine-tuning. The MSSM by
itself does not explain why the $\mu$ parameter should be at the EW
scale. Considering that the motivation of the supersymmetrization of
the SM was to resolve a fine-tuning problem (gauge hierarchy
problem), the fine-tuning problem of the MSSM is a serious issue.

We consider  various extended MSSM models that resolve this
$\mu$-problem and compare their lightest neutralino properties to
that of the MSSM. In these beyond-MSSM models, at least one Higgs
singlet is commonly present and generates an effective $\mu$
parameter when the associated symmetry is broken at the EW or TeV
scale. Because of the superpartners of the Higgs singlets or the
extra gauge boson, the neutralino sector in these models may be
significantly different from that of the MSSM. We investigate the
mass and the coupling of the lightest neutralino of each model
allowed by the model parameters and the experimental data.

In Section \ref{models}, we describe the models and their neutralino
sectors. In Section \ref{direct}, we analyze the lightest
neutralinos allowed by the model parameters and direct experimental
constraints. In Section \ref{indirect}, we discuss indirect
observational constraints on these lightest neutralinos. In Section
\ref{discussion}, we discuss our numerical results, and then
summarize our results in Section \ref{conclusion}.

\section{Models}
\label{models}
\begin{table}[t]
\caption{Higgses and Neutralinos of the MSSM and its extensions
\label{table:model}}
\begin{tabular}{c|c|l|l}
\hline \hline
Model   & Symmetry & ~Higgses (CP even, CP odd, charged) & ~Neutralinos \\
\hline
MSSM    & --                              & $H_1^0, H_2^0, A^0, H^\pm$ & $\tilde B, \tilde W_3, \tilde H_1^0, \tilde H_2^0$ \\
NMSSM   & $\mathbb Z_3$                   & $+~ H_3^0, A_2^0$           & $+~ \tilde S$ \\
nMSSM   & $\mathbb Z^R_5, \mathbb Z^R_7$  & $+~ H_3^0, A_2^0$           & $+~ \tilde S$ \\
UMSSM   & $U(1)'$                         & $+~ H_3^0$                  & $+~ \tilde S, \tilde Z'$ \\
S-model & $U(1)'$                         & $+~ H_3^0, H_4^0, H_5^0, H_6^0, A_2^0, A_3^0, A_4^0$ & $+~ \tilde S, \tilde Z', \tilde S_1, \tilde S_2, \tilde S_3$ \\
\hline \hline
\end{tabular}
\end{table}

The extended MSSM models that we consider are the Next-to-Minimal
Supersymmetric SM (NMSSM) \cite{NMSSM}, the Minimal Non-minimal
Supersymmetric SM (MNSSM) a.k.a. the nearly Minimal Supersymmetric
SM (nMSSM) \cite{nMSSM}, the $U(1)'$-extended Minimal Supersymmetric
SM (UMSSM) \cite{UMSSM}, and the $U(1)'$-extended Supersymmetric SM
with a secluded $U(1)'$-breaking sector (S-model) \cite{S-model}.
All of these extended models prevent the $\mu$-term ($\mu H_1 H_2$)
and allow an effective $\mu$-term ($S H_1 H_2$) through a VEV
$\left< S \right>$ of a Higgs singlet associated with a new symmetry. The
NMSSM and the nMSSM adopt discrete symmetries while the UMSSM and
the S-model use an Abelian gauge symmetry spontaneously broken by
the Higgs singlet.

\subsection{Superpotentials}
The superpotential of Higgses (both isospin doublets and singlets)
for each model is given below\footnote{The term $\alpha S$ in the
nMSSM is a loop-generated tadpole term that breaks the discrete
symmetry; see Ref. \cite{nMSSM}.}. \bea
W_{\rm MSSM} = \mu H_1 H_2 ~~~~~~~~~~~~~~ \qquad \\
W_{\rm NMSSM} = h_s S H_1 H_2 + \frac{\kappa}{3} S^3 \qquad && W_{\rm nMSSM} = h_s S H_1 H_2 + \alpha S \\
W_{\rm UMSSM} = h_s S H_1 H_2 ~~~~~~~~~ \qquad && W_{\rm S-model} =
h_s S H_1 H_2 + \lambda_s S_1 S_2 S_3 \eea The other parts of the
superpotentials are the Yukawa terms of the MSSM and possible extra
terms related to the exotic chiral fields in the $U(1)'$ models
needed to cancel anomalies. The exotic chiral terms are
model-dependent and we do not specify them here, assuming that the
masses of exotic fields are heavy enough to give insignificant
effects to the EW scale phenomenology that we are interested in.
Specific examples of models with exotic fields can be found in Ref.
\cite{UMSSM, S-model}.

The Higgses and neutralinos of each model are listed in Table
\ref{table:model}. In the $U(1)'$-extended model, the addition of
one Higgs singlet does not give an additional CP odd Higgs since a
goldstone boson is absorbed to be the longitudinal mode of the
massive $U(1)'$ gauge boson, $Z'$.

We use the common notation of $h_s$ for the coefficient of the $S
H_1 H_2$ term in each model for easy comparison. In every model the
dynamically generated effective $\mu$ parameter is given by \bea
\mu_{\rm eff} = h_s \left< S \right> \eea and therefore the VEV of
the Higgs singlet or the symmetry breaking scale needs to be at the
EW/TeV scale.

\subsection{Neutralino mass matrices}
The MSSM has 4 neutralinos ($\tilde B$, $\tilde W_3$, $\tilde
H^0_1$, $\tilde H^0_2$) while the NMSSM has 5 neutralinos (MSSM
components + $\tilde S$). The neutralino mass matrix of the NMSSM in
the $\{ \tilde B$, $\tilde W_3$, $\tilde H^0_1$, $\tilde H^0_2$,
$\tilde S \}$ basis is given by \bea M_{\chi^0}= \left( \matrix{ M_1
& 0 & - g_1 v_1 / 2 & g_1 v_2 / 2 & 0 \cr 0 & M_2 & g_2 v_1 / 2 & -
g_2 v_2 / 2 & 0 \cr - g_1 v_1 / 2 & g_2 v_1 / 2 & 0 & - \mu_{\rm
eff} & - \mu_{\rm eff} v_2 / s \cr g_1 v_2 / 2 & - g_2 v_2 / 2 & -
\mu_{\rm eff} & 0 & - \mu_{\rm eff} v_1 / s \cr 0 & 0 & - \mu_{\rm
eff} v_2 / s & - \mu_{\rm eff} v_1 / s & 2 \kappa s / \sqrt{2} }
\right). \label{eqn:massmatrixNMSSM} \eea The first $4 \times 4$
submatrix corresponds to the MSSM limit, which can be obtained from
Eq. (\ref{eqn:massmatrixNMSSM}) by taking $s \gg \cal{O}$(EW).
Depending on the value of the parameter $\kappa$, the 5th component
may be very heavy and decoupled from the other EW scale components
or almost massless. The NMSSM assumes a discrete symmetry ${\mathbb
Z_3}$ to avoid the $\mu$-term, but allows $S^3$ term in the
superpotential. The VEVs of the Higgses are defined as $\langle
H_i^0 \rangle \equiv \frac{v_i}{\sqrt{2}}$ with $\sqrt{v_1^2 +
v_2^2} \equiv v \simeq 246$ GeV and $\langle S \rangle \equiv
\frac{s}{\sqrt{2}}$. The gauge couplings are $g_1 = e /
\cos\theta_W$ and $g_2 = e / \sin\theta_W$. The NMSSM is one of the
simplest extensions of the MSSM, but the ${\mathbb Z_3}$ symmetry
predicts domain walls which are not observed \cite{domainwall}.

The nMSSM \cite{nMSSM} was devised to avoid the domain wall problem
while maintaining a discrete symmetry. The nMSSM has the same 5
neutralinos and the same mass matrix as Eq.
(\ref{eqn:massmatrixNMSSM}) except that the $\kappa$ term in the
$(5,5)$ entry vanishes since it is not allowed by the discrete
symmetry.

The UMSSM uses an Abelian gauge symmetry instead of a discrete
symmetry and thus is free from the domain wall problem. With the
superpartner of the $U(1)'$ gauge boson ($Z'$), the UMSSM has 6
neutralinos (NMSSM or nMSSM components + $\tilde Z'$) and its mass
matrix in the basis of $\{\tilde{B}$, $\tilde{W}_3$,
$\tilde{H}_1^0$, $\tilde{H}_2^0$, $\tilde{S}$, $\tilde{Z'} \}$ is
given by \bea M_{\chi^0}= \left( \matrix{ M_1 & 0 & - g_1 v_1 / 2 &
g_1 v_2 / 2 & 0 & 0 \cr 0 & M_2 & g_2 v_1 / 2 & - g_2 v_2 / 2 & 0 &
0 \cr - g_1 v_1 / 2 & g_2 v_1 / 2 & 0 & - \mu_{\rm eff} & - \mu_{\rm
eff} v_2 / s & g_{Z'} Q'_{H_1} v_1 \cr g_1 v_2 / 2 & - g_2 v_2 / 2 &
- \mu_{\rm eff} & 0 & - \mu_{\rm eff} v_1 / s & g_{Z'} Q'_{H_2} v_2
\cr 0 & 0 & - \mu_{\rm eff} v_2 / s & - \mu_{\rm eff} v_1 / s & 0 &
g_{Z'} Q'_S s \cr 0 & 0 & g_{Z'} Q'_{H_1} v_1 & g_{Z'} Q'_{H_2} v_2
& g_{Z'} Q'_S s & M_{1'} } \right). \label{eqn:massmatrixUMSSM} \eea
The first $5 \times 5$ submatrix corresponds to the nMSSM limit (or
NMSSM limit in the special case $\kappa = 0$) that can be obtained
from Eq. (\ref{eqn:massmatrixUMSSM}) by taking $M_{1'} \gg {\cal
O}$(EW). In this limit the mass of the $Z'$-ino becomes very large
and this component decouples from the others. Here $g_{Z'}$ is the
$U(1)'$ gauge coupling constant and $Q'$ is the $U(1)'$ charge. The
$U(1)'$ charges should satisfy \be Q'_{H_1} + Q'_{H_2} \neq 0 \qquad
Q'_{H_1} + Q'_{H_2} + Q'_S = 0 \ee in order to replace the
$\mu$-term with the effective $\mu$-term dynamically generated by
the Higgs singlet $S$. For the numerical analysis in this paper we
use the $\eta$-model charge assignments and the Grand Unification
Theory (GUT) motivated gauge coupling, $g_{Z'}$: \bea Q'_{H_1} =
\frac{1}{2 \sqrt{15}} \qquad Q'_{H_2} = \frac{4}{2 \sqrt{15}} \qquad
g_{Z'} = \sqrt{\frac{5}{3}} g_1 \eea

The S-model was introduced to resolve tension between the EW scale
$\mu_{\rm eff}$ and the heavy $Z'$ (up to multi-TeV scale). It is
basically the extension of the UMSSM with 3 additional Higgs
singlets to provide additional contributions to the $Z'$ mass while
keeping $\mu_{\rm eff} = h_s \left< S \right>$ at the EW scale. The
S-model has 9 neutralinos (UMSSM components + $\tilde S_1$, $\tilde
S_2$, $\tilde S_3$), and its mass matrix has 3 more columns/rows
added to Eq. (\ref{eqn:massmatrixUMSSM}). (See Ref. \cite{S-model}
for the S-model and its full $9 \times 9$ neutralino mass matrix.)
Its first $6 \times 6$ submatrix corresponds to the UMSSM limit,
which can be realized by taking $s_{1, 2, 3} \gg {\cal O}$(EW) with
$\lambda_s$ comparable to gauge couplings. However, the most
realistic case is for small $\lambda_s$ and large $s_{1, 2, 3}$
\cite{S-model}, in which case four of the neutralinos, consisting
almost entirely of  $\tilde Z'$, $\tilde S_1$, $\tilde S_2$, and
$\tilde S_3$, essentially decouple from the others.
 Since the full $9 \times 9$ matrix has
a number of free parameters, we consider only this decoupling limit
when we discuss the light neutralinos, where there are 5 neutralinos
with masses and compositions the same as the nMSSM\footnote{The four
decoupled neutralinos typically consist of one heavy pair involving
the $\tilde Z'$ and one linear combination of $\tilde S_1$, $\tilde
S_2$, $\tilde S_3$, as well as two more states associated with the
orthogonal combinations of $\tilde S_{1,2,3}$
\cite{S-model,S-model_Higgs}. The latter can be light, or even be
the lightest neutralino in limiting cases. We do not consider that
possibility here.}.

\subsection{Interesting limits for the neutralino masses}
In this section, we analyze the neutralino mass matrix more closely,
especially for the nMSSM and UMSSM limits. The diagonalization of
the neutralino mass matrix is accomplished via a unitary matrix $N$
as \bea N^T M_{\chi^0} N = {\rm Diag}( M_{\chi^0_1}, M_{\chi^0_2},
\cdots ). \eea The singlino ($\tilde S$) composition of the lightest
neutralino ($\chi^0_1$) is $|N_{15}|^2$.

The nMSSM allows the possibility of a very light or massless
neutralino which is mainly singlino. This is apparent from
(\ref{eqn:massmatrixNMSSM}) in the limit of very large $s$ (and
$\kappa=0$), for which one obtains a massless singlino. However, the
tendency for a light singlino-dominated neutralino persists even for
smaller $s$. In fact, it is  possible to have an exactly massless
eigenvalue, which occurs for
 \bea {\rm Det}(M_{\chi^0}) = \left( \frac{\mu v}{s} \right)^2
 \left(M_Z^2 (M_1 \cos^2\theta_W + M_2 \sin^2\theta_W) -
\mu M_1 M_2 \sin2\beta\right) = 0,  \label{detm} \eea which leads to
\bea M_Z^2 \approx 0.8 \mu M_2 \sin2\beta, \label{eqn:massless} \eea
where we have assumed the gaugino mass unification condition $M_1
\simeq 0.5 M_2$. Eq. (\ref{eqn:massless}) can easily be satisfied
with EW/TeV scale values of $\mu$ and $M_2$. It is interesting that
this can occur only for $\mu M_2 > 0$, which is favored by the
deviation of the muon anomalous magnetic moment data from the SM
expectation \cite {g-2_theory} \cite{g-2_exp}, and also by $b \to s
\gamma$ data in most cases \cite {SUGRAWG}. In practice, parameters
satisfying condition (\ref{eqn:massless}) are often excluded by the
chargino mass and $Z$ width constraints discussed in Section
\ref{direct}. Nevertheless, there are allowed points lying nearby in
which the combination of moderate $s$ and the smallness of $M_Z^2 -
0.8 \mu M_2 \sin2\beta$ lead to a very light singlino-dominated
neutralino. For large enough $s$ the neutralino can be massless. We
will further discuss the massless neutralino scenario later.

Now we consider interesting limits of the UMSSM.

(i) In the case that only $M_{1'}$, the diagonal element of the 6th
row/column of Eq. (\ref{eqn:massmatrixUMSSM}), is very large
compared to other mass parameters, the $\tilde Z'$ will be very
heavy and decoupled, leaving a mass matrix similar to that of the
nMSSM.

(ii) When only $s$ is very large compared to other mass parameters
(with $\mu_{\rm eff}$ at the EW scale, which requires $h_s\sim 0$),
the elements of the 5th row/column in Eq.
(\ref{eqn:massmatrixUMSSM}) are small except for the 6th component.
Then the effective mass matrix for the 5th and the 6th components is
\bea M_{\chi^0 [5,6]} = \left( \matrix{ 0 & g_{Z'} Q'_S s \cr g_{Z'}
Q'_S s & 0} \right), \label{eqn:massmatrix56a} \eea which gives
almost degenerate physical masses of $g_{Z'} |Q'_S| s$. These two
heavy states approximately decouple from the other four neutralinos,
which are MSSM-like.

(iii) When both $M_{1'}$ and $s$ are large compared to other mass
parameters and $\mu_{\rm eff}$ is at the EW scale (i.e., $h_s$ is
very small), the effective mass matrix for the 5th and 6th
components is \bea M_{\chi^0 [5,6]} = \left( \matrix{ 0 & g_{Z'}
Q'_S s \cr g_{Z'} Q'_S s & M_{1'}} \right) \label{eqn:massmatrix56b}
\eea which has eigenvalues \bea \frac{1 - \sqrt{1 + 4 \rho^2}}{2}
M_{1'} \quad \mbox{and} \quad \frac{1 + \sqrt{1 + 4 \rho^2}}{2}
M_{1'}\label{eqn:mmeigns} \eea where $\rho \equiv |g_{Z'} Q'_S s /
M_{1'}|$. The corresponding neutralino masses in various limits are
\bea
0, \quad |M_{1'}| \quad (\rho \ll 1) \label{eqn:masslesslimit} \\
\rho |M_{1'}|, \quad \rho |M_{1'}| \quad (\rho \gg 1) \\
0.62 |M_{1'}|, \quad 1.62 |M_{1'}| \quad (\rho = 1) \eea The $\rho
\ll 1$ case corresponds to taking the UMSSM to the nMSSM limit
($M_{1'} \gg \cal{O}$(EW)) and then to the MSSM limit ($s \gg
\cal{O}$(EW)) consecutively. Then a very light neutralino state
along with a very heavy neutralino state occurs. Besides these
lightest (dominated by singlino) and heaviest (dominated by
$Z'$-ino) states, the other neutralinos are basically the same as
the MSSM neutralinos. The $\rho \gg 1$ case is similar to the Eq.
(\ref{eqn:massmatrix56a}) case. For the range of $\rho = 10 \sim
0.1$, the mass ratio of the lighter state to the heavier one is $1
\sim 100$.

(iv) The case of gaugino mass unification $M_{1'} = M_1 =
\frac{5}{3} \frac{g_1^2}{g_2^2} M_2 \simeq 0.5 M_2$ will be
considered in the next section.

\section{Direct Constraints on the Lightest Neutralinos}
\label{direct}
Here we discuss the allowed mass range of the lightest
neutralino after incorporating direct constraints from the LEP
experiments and the model parameter structure.

\subsection{LEP constraints on the light chargino mass and the $Z$ width }
All the models considered have only one charged gaugino ($\tilde
W^\pm$) and one charged Higgsino ($\tilde H^\pm$) and thus have a
common chargino mass matrix which is the same as in the MSSM, \bea
M_{\chi^\pm} = \left( \matrix{M_2 & \sqrt{2} M_W \sin\beta \cr
\sqrt{2} M_W \cos\beta & \mu_{\rm eff}} \right). \eea The LEP2 data
requires the light chargino mass to be $M_{\chi_1^\pm} > 104$ GeV
\cite{charginoMass} which gives lower bounds on $M_2$ and $\mu_{\rm
eff}$ for a fixed $\tan\beta$.

For a lightest neutralino with $M_{\chi^0_1} \le M_Z / 2$, the LEP
constraint $\Gamma_Z^{\rm exp} - \Gamma_Z^{\rm SM} = (-2.0 \pm 2.6)$
MeV \cite{DeltaZwidth} must be satisfied. Since the $Z$ boson does
not couple directly to the singlino or $Z'$-ino, the
$Z$-$\chi_1^0$-$\chi_1^0$ coupling for every model is the same as
that of the MSSM. The expression for the partial width for $Z$ decay
to a pair of the lightest neutralinos is \bea \Gamma_{Z \to \chi_1^0
\chi_1^0} = \frac{g_1^2 + g_2^2}{4 \pi} \frac{ (|N_{13}|^2 -
|N_{14}|^2)^2 }{24 M_Z^2} \left( M_Z^2 - (2 M_{\chi_1^0})^2
\right)^{3/2} \Theta(M_Z - 2 M_{\chi^0_1}). \label{eqn:Zwidth} \eea
For a lightest neutralino of very small mass ($M_{\chi^0_1} \ll M_Z
/ 2$) to be allowed\footnote{There are some points involving two or
more very light neutralinos, for which the lightest neutralino
satisfies the $Z$ width constraint but the others do not. This
generally occurs for small $M_2$ and $\mu$ values which are already
excluded by the chargino mass constraint.}, its two doublet Higgsino
compositions should be either almost identical ($|N_{13}|^2 \approx
|N_{14}|^2$) or negligible ($|N_{13}|^2, |N_{14}|^2 \ll 1$) by
gaugino or singlino dominance.

\subsection{Lightest neutralino mass range allowed by direct constraints}
\begin{table*}
\caption{The lower (upper) bounds on the lightest neutralino mass
$(M_{\chi^0_1})$, the singlino composition $(|N_{15}|^2)$ and the
parameter values at the bounds in various Supersymmetric models for
several fixed $\tan\beta$ values. All mass units are in GeV, except
for $\Gamma_{Z \to \chi_1^0 \chi_1^0}$ in MeV. For the UMSSM, the
$\eta$-model charge assignment is assumed. We consider the range
$0.1 \le h_s \le 0.75$ and only cases where the direct experimental
constraints $M_{\chi^\pm_1} > 104$ GeV and $\Gamma_{Z \to \chi_1^0
\chi_1^0} < 2.3$ MeV ($95\%$ C.L.) are satisfied. The scans are made
for $M_2$, $\mu = 50 \sim 500$ GeV and $s = 50 \sim 2000$ GeV. For
the NMSSM, the parameter range $|\kappa| = 0.1 \sim 0.75$ is
scanned, and the $\kappa > 0$ and $\kappa < 0$ cases are listed
separately. \label{table:mass}}
\begin{ruledtabular}
\tiny
\begin{tabular}{r|rrr|rrrr|rrr}
$\tan\beta$ & Model & $M_{\chi^0_1}~$             & $|N_{15}|^2~~$ & $\mu~~~~~$ & $M_2~~~~$   & $s~~~~~~$    & $\kappa~~~~~~$ & $h_s~~~~~$   & $M_{\chi^\pm_1}~~$ & $\Gamma_{Z \to \chi_1^0 \chi_1^0}$ \\
\hline
$1~~  $     & MSSM         & $55(242)$ &              & $481(500)$ & $121(500)$ &              &                &              & $104(420)$ & ~--  ~(~~--~~) \\
            & NMSSM        & $55(242)$ & $0.00(0.00)$ & $481(500)$ & $121(500)$ & $1850(1605)$ & $0.15(0.30)$   & $0.37(0.44)$ & $104(420)$ & ~--  ~(~~--~~) \\
            &              & $0(242)$  & $0.78(0.00)$ & $242(500)$ & $265(500)$ & $470(1995)$  & $-0.10(-0.40)$ & $0.73(0.35)$ & $173(420)$ & $0.00$(~~--~~) \\
            & nMSSM        & $2(83)$   & $0.99(0.71)$ & $142(121)$ & $486(500)$ & $2000(230)$  &                & $0.10(0.74)$ & $124(105)$ & $0.00$(~~--~~) \\
            & UMSSM        & $0(242)$  & $0.60(0.00)$ & $171(500)$ & $444(500)$ & $365(2000)$  &                & $0.66(0.35)$ & $149(420)$ & $0.55$(~~--~~) \\
\hline
$2~~  $     & MSSM         & $55(243)$ &              & $486(500)$ & $118(500)$ &              &                &              & $104(425)$ & ~--  ~(~~--~~) \\
            & NMSSM        & $48(243)$ & $0.23(0.00)$ & $168(500)$ & $190(500)$ & $320(2000)$  & $0.10(0.65)$   & $0.74(0.35)$ & $104(425)$ & ~--  ~(~~--~~) \\
            &              & $17(243)$ & $0.78(0.00)$ & $255(500)$ & $500(500)$ & $495(1030)$  & $-0.10(-0.20)$ & $0.73(0.69)$ & $234(425)$ & $2.29$(~~--~~) \\
            & nMSSM        & $1(50)$   & $0.98(0.58)$ & $142(151)$ & $248(500)$ & $2000(285)$  &                & $0.10(0.75)$ & $104(136)$ & $0.03$(~~--~~) \\
            & UMSSM        & $26(243)$ & $0.61(0.00)$ & $228(500)$ & $487(500)$ & $430(1200)$  &                & $0.75(0.59)$ & $208(425)$ & $2.29$(~~--~~) \\
\hline
$10~~ $     & MSSM         & $54(247)$ &              & $468(500)$ & $110(500)$ &              &                &              & $104(441)$ & ~--  ~(~~--~~) \\
            & NMSSM        & $21(247)$ & $0.43(0.00)$ & $202(500)$ & $133(500)$ & $405(2000)$  & $0.10(0.65)$   & $0.71(0.35)$ & $104(441)$ & $2.29$(~~--~~) \\
            &              & $37(247)$ & $0.71(0.00)$ & $216(500)$ & $476(500)$ & $415(980)$   & $-0.10(-0.20)$ &$0.74(0.72)$  & $206(441)$ & $2.30$(~~--~~) \\
            & nMSSM        & $0(6)   $ & $0.98(0.87)$ & $161(355)$ & $321(500)$ & $2000(670)$  &                & $0.11(0.75)$ & $145(335)$ & $0.04(2.26)$   \\
            & UMSSM        & $39(247)$ & $0.55(0.00)$ & $173(500)$ & $499(500)$ & $330(1180)$  &                & $0.74(0.60)$ & $165(441)$ & $2.30$(~~--~~) \\
\hline
$50~~ $     & MSSM         & $53(248)$ &              & $451(500)$ & $108(500)$ &              &                &              & $104(445)$ & ~--  ~(~~--~~) \\
            & NMSSM        & $16(248)$ & $0.48(0.00)$ & $181(500)$ & $133(500)$ & $405(2000)$  & $0.10(0.65)$   & $0.63(0.35)$ & $104(445)$ & $2.29$(~~--~~) \\
            &              & $39(248)$ & $0.66(0.00)$ & $191(500)$ & $475(500)$ & $365(980)$   & $-0.10(-0.20)$ &$0.74(0.72)$  & $184(445)$ & $2.30$(~~--~~) \\
            & nMSSM        & $0(10)  $ & $0.98(0.76)$ & $500(246)$ & $500(118)$ & $2000(590)$  &                & $0.35(0.59)$ & $445(104)$ & $0.04(2.29)$   \\
            & UMSSM        & $41(248)$ & $0.50(0.00)$ & $147(500)$ & $500(500)$ & $280(1170)$  &                & $0.74(0.60)$ & $143(445)$ & $2.30$(~~--~~)
\end{tabular}
\end{ruledtabular}
\end{table*}

We evaluate the bounds on the lightest neutralino mass
($M_{\chi^0_1}$) and the singlino component of $\chi^0_1$ in the
MSSM, the NMSSM, the nMSSM, and the UMSSM with gaugino mass
unification. The nMSSM also represents the NMSSM in the $\kappa \to
0$ limit, the UMSSM in the $M_{1'} \gg M_1$ limit, and the
decoupling limit of the S-model.

We require that the tree-level masses satisfy the direct LEP limits
of $M_{\chi^\pm_1} > 104$ GeV, and $\Gamma_{Z \to \chi_1^0 \chi_1^0}
< 2.3$ MeV ($95\%$ C.L.). Bounds from naturalness and perturbativity
constraints \cite{S-model,Miller:2003ay,neutralino_nMSSM} are also
imposed on the couplings of $0.1 \le h_s \le 0.75$ and $\sqrt{h_s^2
+ \kappa^2} \le 0.75$ (for the NMSSM)\footnote{An exact $h_s$ bound
(and its source) may be a little different depending on models, but
we assume a common bound for easy comparison. A lower bound on the
NMSSM $|\kappa|$ is fuzzy except that $\kappa \ne 0$ should be
satisfied to avoid an unacceptable Peccei-Quinn symmetry. We set
$|\kappa| \ge 0.1$ as a lower bound; results for a smaller
$|\kappa|$ can be described by an interpolation of the nMSSM result
($\kappa \to 0$ limit).}. The LEP2 Higgs mass bound of $m_h > 114$
GeV does not apply directly to the extended MSSM models where the
physical Higgses exist as mixtures of doublets and singlets
\cite{mixedhiggs}.

We choose a phase convention in which $\mu$ and the VEVs are real
and positive. $\kappa$ and the gaugino masses can in principle be
complex, but we restrict our considerations to real values. We
mainly consider the favored case of positive gaugino masses, but
comment on the effects of negative masses. We scan $M_2$, $\mu = 50
\sim 500$ GeV with a step size of $1$ GeV\footnote{We exclude very
small $M_2$ or $\mu$ to avoid very light non-singlino states. These
are also excluded by the chargino mass constraints.}, $s = 50 \sim
2000$ GeV with a step size of $5$ GeV, $\tan\beta = 0.5$, $1$,
$1.5$, $2$, $10$, $50$, and $|\kappa| = 0.1 \sim 0.75$ with a step
size of $0.05$ (for the NMSSM). The gaugino mass unification
relation $M_{1'} = M_1 = \frac{5}{3} \frac{g_1^2}{g_2^2} M_2 \simeq
0.5 M_2$ is assumed. The Higgs singlet VEV $s$ cannot be too small
in the UMSSM because it is responsible for the mass of the $Z'$, but
there are some possible ways to get around this, as we will discuss
in Section \ref{sec:Z'mass}.

The bounds that we obtain on the the lightest neutralino mass are
\bea
53 \mbox{ GeV} &\le& M_{\chi^0_1}~ \le~ 248 \mbox{ GeV  \quad [MSSM]} \label{eqn:MSSMrange} \\
0  \mbox{ GeV} &\le& M_{\chi^0_1}~ \le~ 248 \mbox{ GeV  \quad [NMSSM]} \\
0  \mbox{ GeV} &\le& M_{\chi^0_1}~ \le~ {\tiny~} ~83 \mbox{ GeV
\quad [nMSSM, S-model (decoupling limit)]} \\
0  \mbox{ GeV} &\le& M_{\chi^0_1}~ \le~ 248 \mbox{ GeV  \quad
[UMSSM]} \label{eqn:UMSSMrange} \eea Table \ref{table:mass} shows
the lightest neutralino mass ($M_{\chi^0_1}$) and its singlino
composition ($|N_{15}|^2$) as well as other relevant parameter
values at the bounds. The results in the table are only for positive
gaugino masses, but negative gaugino masses do not change these
ranges significantly\footnote{With $M_2 = -50 \sim -500$ GeV, the
$\chi^0_1$ mass ranges are $M_{\chi^0_1} = 39 \sim 254$ GeV (MSSM),
$M_{\chi^0_1} = 0 \sim 254$ GeV (NMSSM), $M_{\chi^0_1} = 0.4 \sim
96$ GeV (nMSSM), $M_{\chi^0_1} = 39 \sim 254$ GeV (UMSSM).}.

Figure \ref{fig:tanb} shows the $M_{\chi^0_1}$ dependence on
$\tan\beta$ and $|N_{15}|^2$ (and also $|N_{16}|^2$ for the UMSSM).
The dashed lines are the MSSM bounds which neglect the rather weak
$\tan\beta$ dependence and use the values of Eq.
(\ref{eqn:MSSMrange}). The mass bound dependence on $\tan\beta$ is
quite sensitive in some models. For example, at $\tan\beta \approx
1$, the NMSSM and the UMSSM have massless $\chi^0_1$ while the nMSSM
has an upper $M_{\chi^0_1}$ bound; the MSSM violates the LEP2 $m_h$
constraint at this $\tan\beta$. Singlino dominance is typical in the
$\chi^0_1$ of the extended MSSM models. Especially, when the
$M_{\chi^0_1}$ is much smaller than the MSSM lower limit
($M_{\chi^0_1} \sim 50$ GeV), the singlino is always the dominant
component. We will further discuss the allowed mass ranges along
with additional indirect constraints in Section \ref{discussion}.

\section{Indirect Constraints on the Lightest Neutralinos}
\label{indirect}
\subsection{Cold dark matter relic density and muon anomalous magnetic moment}
The cold dark matter density is tightly constrained by the WMAP
(CMB) and SDSS (Large Scale Structure) data \cite{relic_exp} to be
(with $1 \sigma$ uncertainty) \bea \Omega_{\rm CDM} h^2 = 0.12 \pm
0.01 \qquad \mbox({\rm WMAP+SDSS)} \label{eqn:relic} \eea where $h =
0.72 \pm 0.08$ is the present day Hubble constant $H_0$ in units of
$100$ km s$^{-1}$ Mpc$^{-1}$ \cite{hubble}. We assume that
$\chi^0_1$ is the sole dark matter and impose the relic density
constraint\footnote{It was emphasized in \cite{nonthermal} that
predicted values of $ \Omega_{\rm CDM} h^2$ lower than the observed
range in (\ref{eqn:relic}) may be allowed if there are nonthermal
production mechanisms.} of Eq. (\ref{eqn:relic}).

\begin{figure}[t]
\begin{minipage}[b]{1\textwidth}
\begin{minipage}[b]{.49\textwidth}
\centering\leavevmode
\epsfxsize=3in
\epsfbox{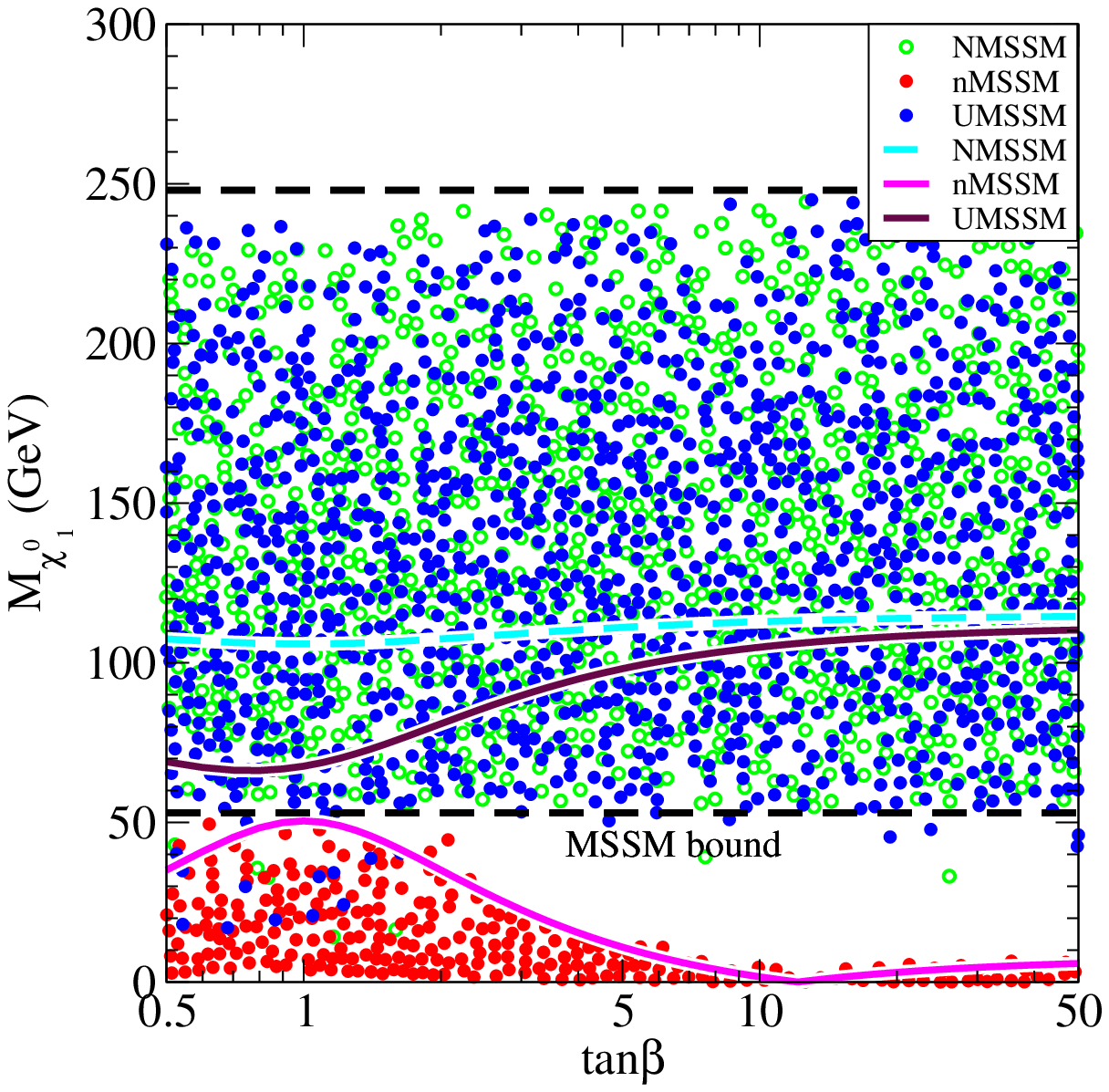}
\end{minipage}
\hfill
\begin{minipage}[b]{.49\textwidth}
\centering\leavevmode \epsfxsize=3in
\epsfbox{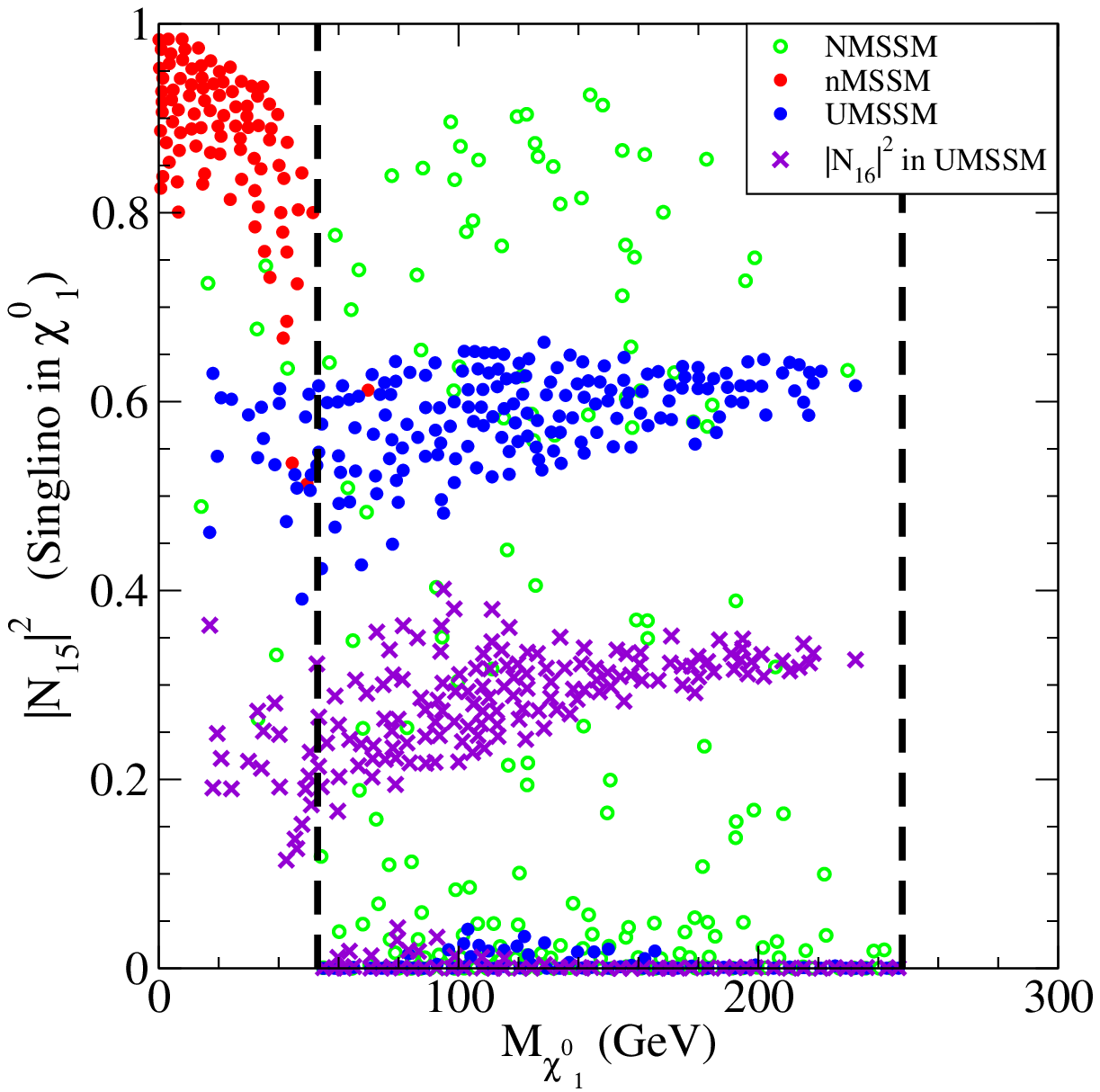}
\end{minipage}
(a)\hspace{0.48\textwidth}(b)
\end{minipage}
\caption{Scatter plots of (a) the $\chi^0_1$ mass ($M_{\chi_1^0}$)
versus $\tan\beta$ and (b) the $\chi^0_1$ singlino composition
($|N_{15}|^2$) versus $M_{\chi_1^0}$ for various models. The crosses
in (b) represent the $Z'$-ino composition ($|N_{16}|^2$) for the
UMSSM. The direct constraints of Section \ref{direct} are imposed.
The solid curves represent a fixed set of inputs $M_2 = 250$ GeV,
$\mu = 250$ GeV, $s = 500$ GeV and $\kappa = 0.5$ (for the NMSSM)
without the direct constraints. The upper (lower) UMSSM singlino
band in (b) corresponds approximately to moderate (large) value of
$s$, with the lightest neutralino being MSSM-like for large $s$.}
\label{fig:tanb}
\end{figure}

\begin{figure}[t]
\begin{minipage}[b]{1\textwidth}
\begin{minipage}[b]{.49\textwidth}
\centering\leavevmode
\epsfxsize=3in
\epsfbox{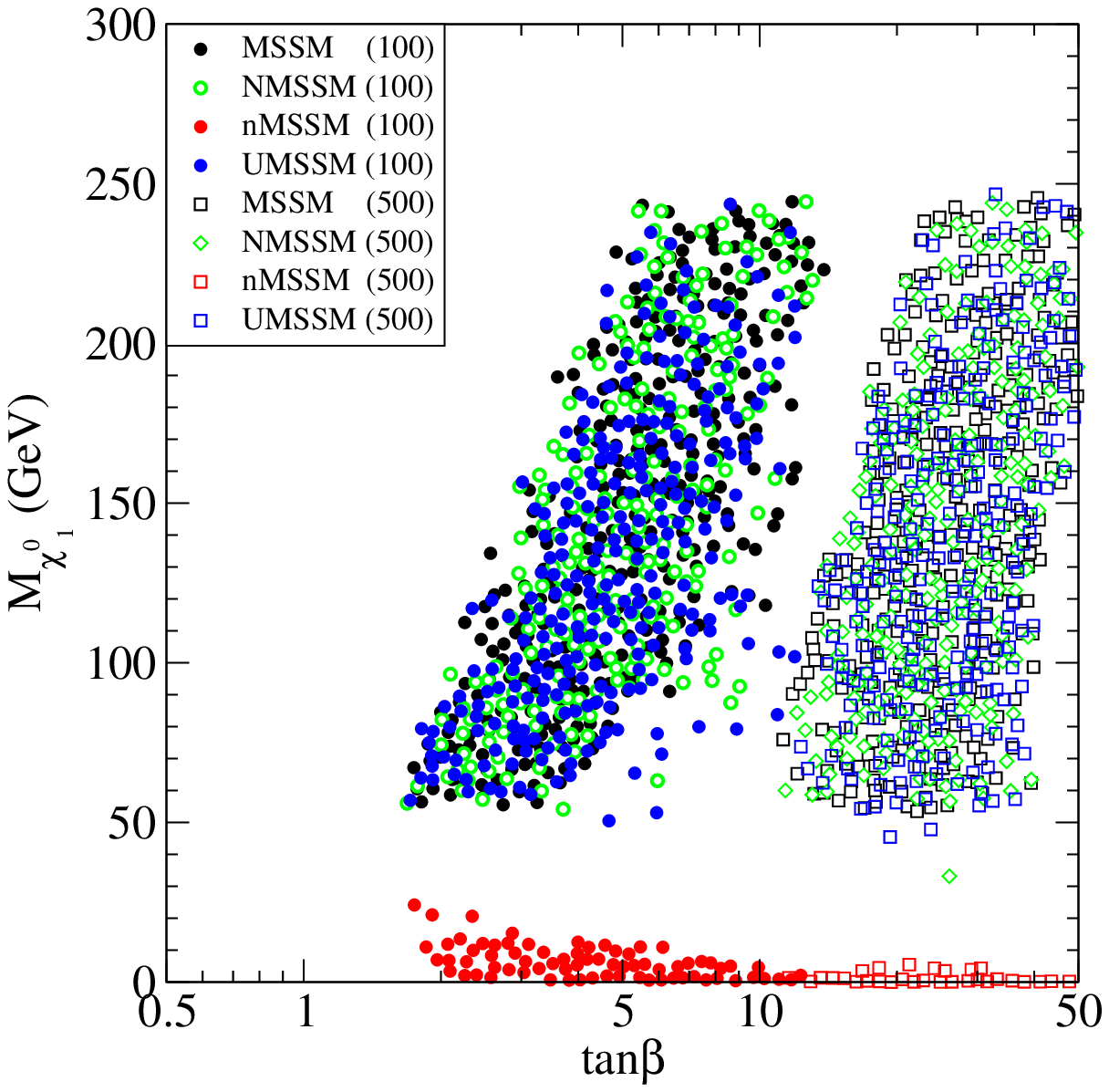}
\end{minipage}
\hfill
\begin{minipage}[b]{.49\textwidth}
\centering\leavevmode
\epsfxsize=3in
\epsfbox{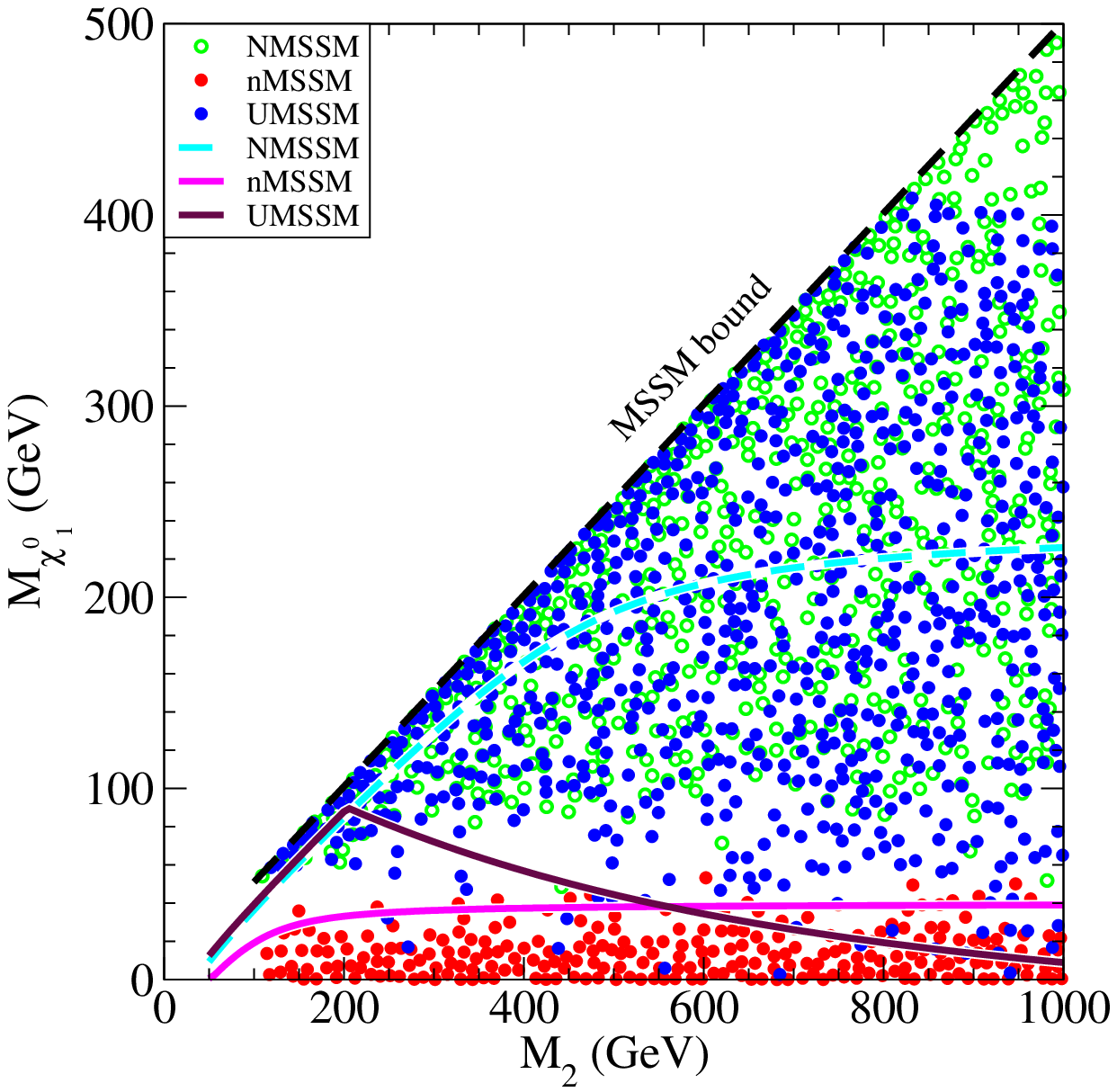}
\end{minipage}
(a)\hspace{0.48\textwidth}(b)
\end{minipage}
\caption{(a) Similar plot as Figure \ref{fig:tanb}(a) that satisfy
the $2.4\sigma$ allowed $(g-2)_\mu$ measurement with $m_L = m_E =
100$ GeV (dots) and $500$ (GeV) (squares). (b) $M_{\chi^0_1}$ versus
$M_2$ with extended ranges of $M_2$ and $\mu$ up to $1000$ GeV. The
solid curves are for $\mu = 250$ GeV, $s = 500$ GeV, $\tan\beta = 2$
and $\kappa = 0.5$.} \label{fig:M2}
\end{figure}

Another important experimental result that can constrain new physics
models is the BNL E821 measurement of the muon anomalous magnetic
moment $a_\mu \equiv (g-2)_\mu / 2$ \cite{g-2_exp}. The deviation
from the SM is \bea \Delta a_\mu \equiv a_\mu({\rm exp}) -
a_\mu({\rm SM}) = (23.9 \pm 10.0) \times 10^{-10} \eea when the SM
prediction is based on the hadronic contribution from $e^+ e^-$
data. However, the $2.4 \sigma$ deviation is reduced to $0.9 \sigma$
if indirect hadronic $\tau$ decay data are used instead for the SM
prediction.

The dominant Supersymmetric contributions to $(g-2)_\mu$ come from
the chargino-sneutrino and the neutralino-smuon loops. The $\Delta
a_\mu$ result practically constrains the sign of $M_2$ to be
positive in our sign convention of $\mu_{\rm eff} > 0$. The
predicted $(g-2)_\mu$ value sensitively depends on $\tan\beta$ and
the scalar muon mass. The muon trilinear scalar coupling dependence
is neglected in our calculation of $a_\mu$.

Typically, a large $\tan\beta$ value is favored to explain the
sizable $2.4 \sigma$ deviation, but with a rather small scalar muon
mass, quite small $\tan\beta$ values are also acceptable. Figure
\ref{fig:M2}(a) shows how the acceptable parameter ranges from
Figure \ref{fig:tanb}(a) change due to the $\Delta a_\mu$ constraint
for two choices of scalar muon mass, $m_L = m_E = 100$ GeV (dots)
and $500$ GeV (squares).

\begin{figure}[t]
\begin{minipage}[b]{1\textwidth}
\begin{minipage}[b]{.49\textwidth}
\centering\leavevmode
\epsfxsize=3in
\epsfbox{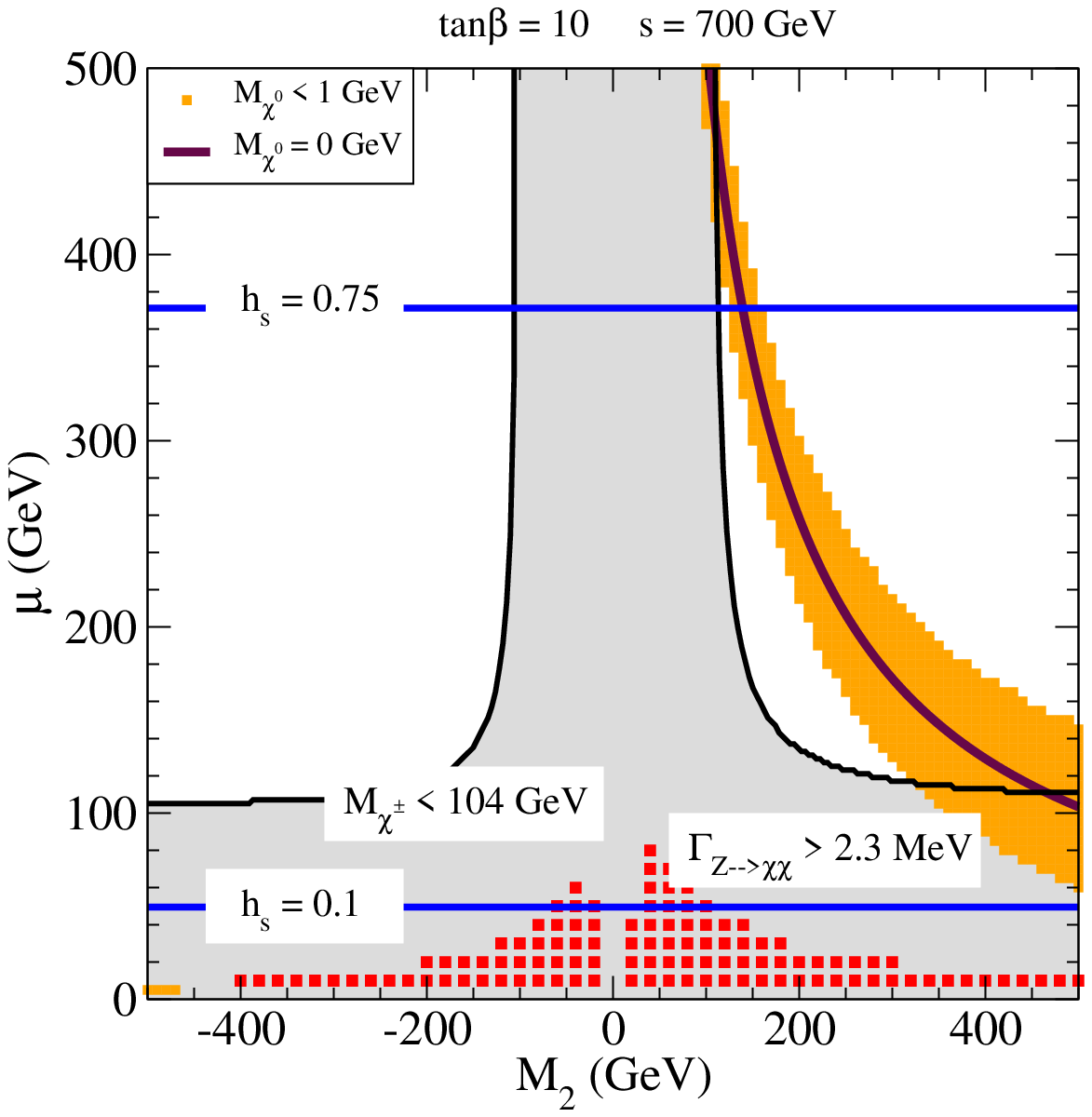}
\end{minipage}
\hfill
\begin{minipage}[b]{.49\textwidth}
\centering\leavevmode
\epsfxsize=3in
\epsfbox{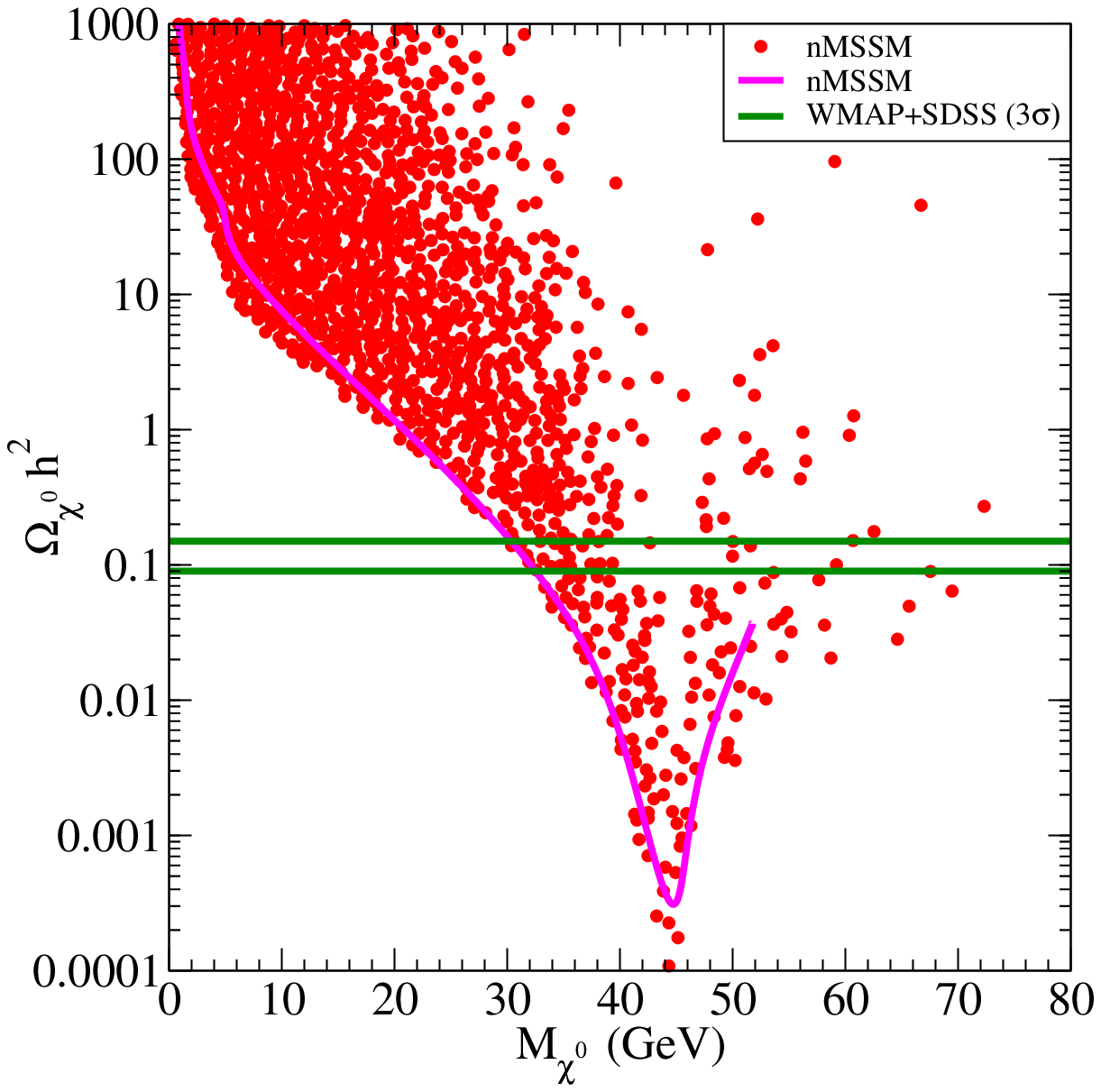}
\end{minipage}
(a)\hspace{0.48\textwidth}(b)
\end{minipage}
\caption{(a) Illustration of the massless neutralinos (dark curve)
and nearby light neutralinos ($M_{\chi^0_1} < 1$ GeV) (orange
region) in the $M_2$-$\mu$ plane along with constraints from
$\Gamma_Z$, $M_{\chi^\pm_1}$ and $h_s$ in the nMSSM or the S-model
in the decoupling limit. Here $\tan\beta = 10$ and $s = 700$ GeV are
assumed. The gap in the $\Gamma_Z$ exclusion region is due to the
emergence of a very light gaugino for $M_{1, 2} \sim 0$. (b) The
relic density for the nMSSM or the S-model with Z-pole annihilation.
The solid curve is for a fixed set of values of $\mu = 200$ GeV,
$s=400$ GeV and $\tan\beta = 1.5$ with $M_2$ varying over $0 \sim
500$ GeV.} \label{fig:relic}
\end{figure}

\subsection{$Z'$ boson mass}
\label{sec:Z'mass}
In $U(1)'$-extended MSSM models, the mass of the new gauge boson
$Z'$ is also an important constraint. The Tevatron Run2 dilepton
data places a lower bound on the $Z'$ mass of $500 \sim 800$ GeV,
with the exact bound depending on the model \cite{ZprimeMass}. In
the UMSSM, the $Z'$ mass is given by \be M_{Z'}^2 = g_{Z'}^2 \left(
Q'^2_{H_1} v_1^2 + Q'^2_{H_2} v_2^2 + Q'^2_S s^2 \right). \ee
Accordingly, the value of $s$ should be at the few TeV level to
satisfy the experimental $M_{Z'}$ bound; this high value of $s$,
with $\mu_{\rm eff} \sim {\cal O}$(EW), requires very small $h_s$
which implies a fine-tuning.

The bound on $s$ can be significantly reduced if there are
additional contributions to $M_{Z'}$. In the S-model, for example,
the $Z'$ mass gets additional contributions from 3 more Higgs
singlet VEVs $s_{1,2,3}$, which practically removes any lower bound
on $s$ if the $s_i$'s are at the TeV-scale. \be M_{Z'}^2 = g_{Z'}^2
\left( Q'^2_{H_1} v_1^2 + Q'^2_{H_2} v_2^2 + Q'^2_S s^2 +
\mbox{$\sum_{i=1}^{3}$} Q'^2_{S_i} s_i^2 \right) \ee These multiple
singlets help to keep $\mu_{\rm eff} = h_s \frac{s}{\sqrt{2}}$ at
the EW scale (resolving the $\mu$-problem) without fine-tuning even
for a multi-TeV scale $Z'$. It is also possible that $Z'$ could be
leptophobic, in which case the Tevatron bound on $M_{Z'}$ is greatly
reduced. Then $s$ is not severely bounded even if there are no
additional contributions to $M_{Z'}$.

\begin{figure}[t]
\begin{minipage}[b]{1\textwidth}
\begin{minipage}[b]{.49\textwidth}
\centering\leavevmode
\epsfxsize=3in
\epsfbox{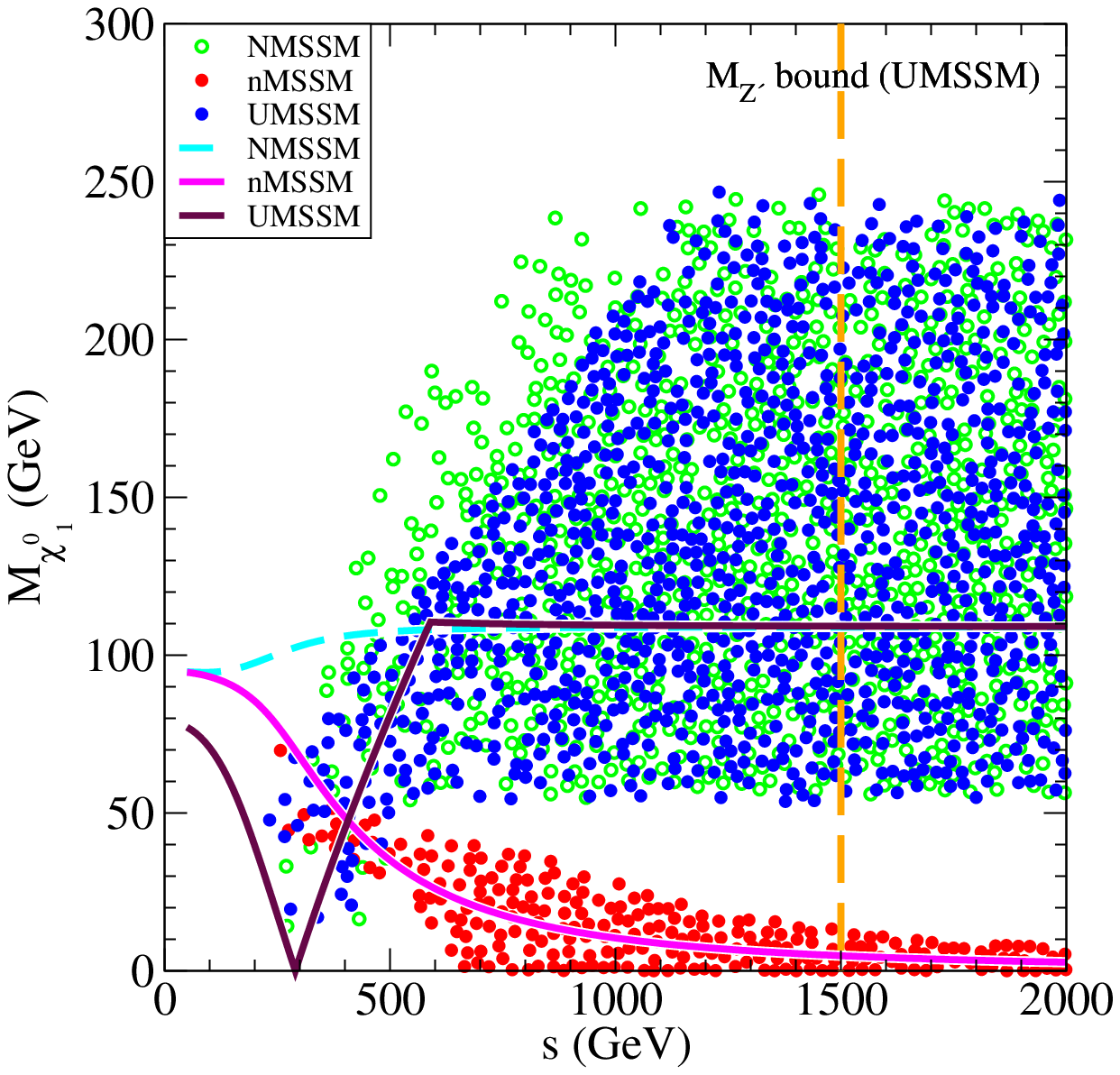}
\end{minipage}
\hfill
\begin{minipage}[b]{.49\textwidth}
\centering\leavevmode
\epsfxsize=3in
\epsfbox{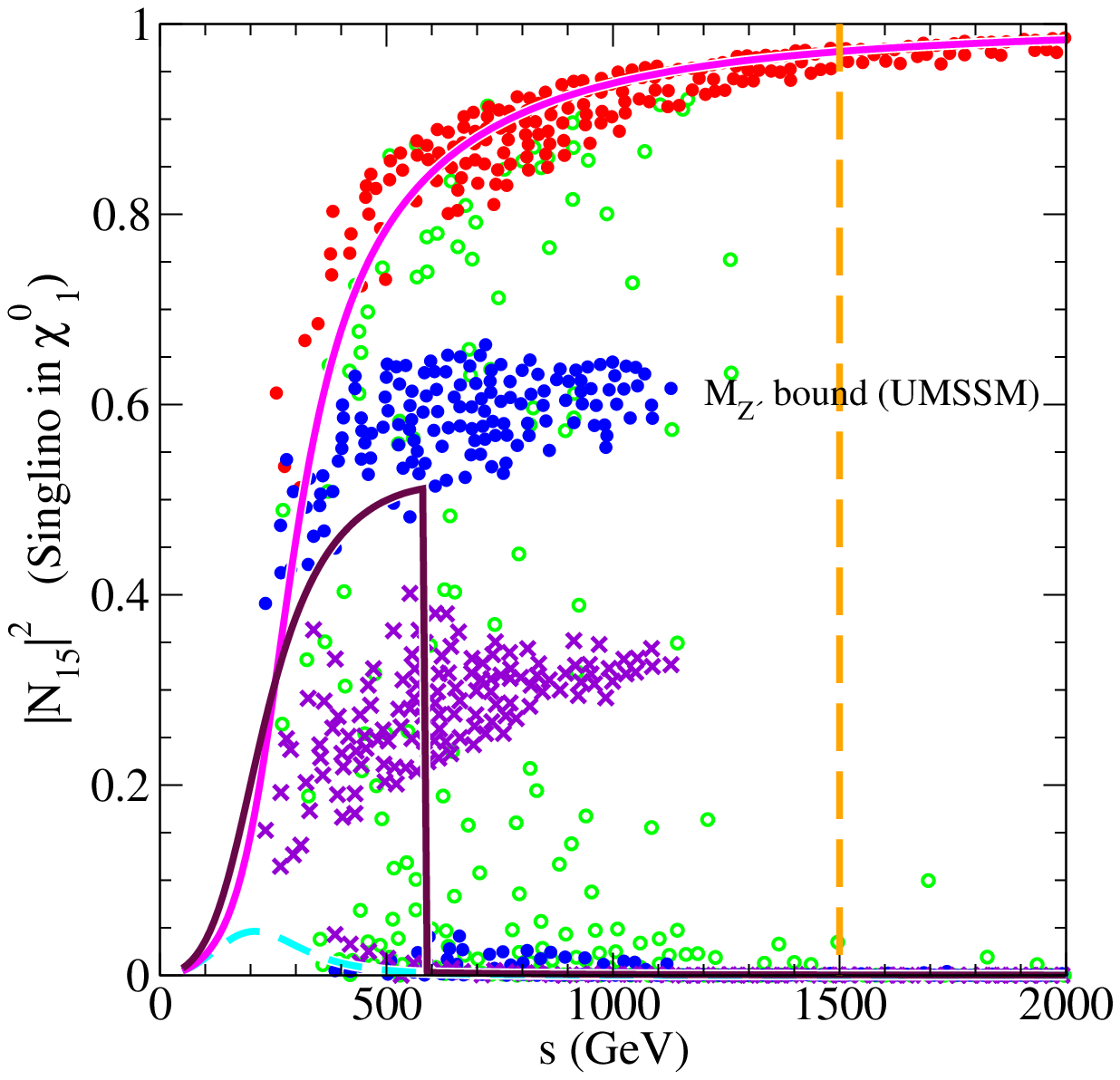}
\end{minipage}
(a)\hspace{0.48\textwidth}(b)
\end{minipage}
\caption{Scatter plots of (a) $M_{\chi^0_1}$ and (b) $|N_{15}|^2$ versus $s$. The solid curves are for $M_2 =
250$ GeV, $\mu = 250$ GeV, $\tan\beta = 2$ and $\kappa = 0.5$. The $M_{Z'}$ bound $s> 1500$ GeV
is approximate and can be evaded, as discussed in Section \ref{sec:Z'mass}.
The crosses in (b) are the $Z'$-ino composition ($|N_{16}|^2$) in the UMSSM.}
\label{fig:s}
\end{figure}

\section{Discussion of the lightest neutralino}
\label{discussion}
\subsection{In the MSSM}
The lightest neutralino mass bound in the MSSM for various
conditions has been well studied \cite{MSSMmass}. In our parameter
range, the upper bound is  $248$ GeV, which is associated with the
assumed upper bound on $M_1$. The $\chi^0_1$ mass bound increases if
the $M_1$ bound increases (Figure \ref{fig:M2}(b)). The lower bound
on $M_{\chi^0_1}$ is determined by the light chargino mass bound of
$M_{\chi^\pm_1} > 104$ GeV; the bound is $M_{\chi^0_1} > 53$ GeV
(roughly half of the chargino mass bound), for which the $\Gamma_{Z
\to \chi^0_1 \chi^0_1}$ constraint is irrelevant. In this
$M_{\chi^0_1}$ range the $\chi^0_1 \chi^0_1$ annihilation cross
section is of suitable size for the general MSSM to provide an
acceptable relic density through multiple annihilation channels
\cite{MSSMCDM}. The $(g-2)_\mu$ deviation can be simultaneously
explained without difficulty \cite{MSSMg-2relic}. The LEP2 SM-like
Higgs mass bound of $m_h> 114$ GeV excludes $\tan\beta \sim 1$ in
the MSSM, though it does not apply to other models with a Higgs
singlet.

\subsection{In the NMSSM}
Systematic studies of the NMSSM neutralinos can be found in Ref.
\cite{chi_NMSSM}. The upper bound on the $\chi^0_1$ mass is $248$
GeV, as in the MSSM. At this point the singlino component is absent
and $\kappa$ is rather large, $0.65$ for $\kappa > 0$ ($-0.20$ for
$\kappa < 0$). With a large enough $\kappa s$ value the singlino
component can be very heavy and decoupled, leaving the rest of the
matrix similar to the MSSM.

With positive $\kappa$, the lower $M_{\chi^0_1}$ bound is $16$ GeV.
With negative $\kappa$, a massless $\chi^0_1$ occurs at $\tan\beta =
1$ with a singlino composition of $|N_{15}|^2$ of $0.78$. Since the
massless $\chi^0_1$ state occurs as the result of mass matrix
mixing, loop effects such as threshold corrections on the gaugino
masses do not prevent the appearance of the massless state. Its
mixing with the right-handed neutrinos may result in interesting
neutrino physics \cite{neutrino}.

This model is disfavored by the non-observation of cosmological
domain walls predicted by the discrete symmetries of the model
\cite{domainwall}. However, there is an approach to interpret the
domain wall as the dark energy \cite{domainwalldarkenergy}. The
domain wall network has to be strongly frustrated (nearly static) to
satisfy the CMB isotropy constraint. In that case the equation of
state of dark energy is expected to be close to $w = -2/3$, which is
disfavored by SNIa data and also by joint analysis with other
cosmological data \cite{domainwalldisfavored} but is not excluded.

\subsection{In the nMSSM or the S-model}
In the nMSSM (also in the S-model in the decoupling limit), a
vanishing lower $\chi^0_1$ mass bound occurs with dominant singlino
composition $|N_{15}|^2 \sim 1$, where $\Gamma_{Z \to \chi^0_1
\chi^0_1}$ of Eq. (\ref{eqn:Zwidth}) is negligible. As discussed
after Eq. (\ref{eqn:massless}), the condition for a massless
neutralino can be easily satisfied regardless of the $\tan\beta$ and
$s$ choices. However, the massless condition for small $\tan\beta$
violates the $M_{\chi_1^\pm} > 104$ GeV constraint. This
 is why the lower bound is not exactly zero for small
$\tan\beta$ in Table \ref{table:mass}, but can nevertheless be very
small for large $s$. Large $s$ is also needed for a light $\chi_1^0$
to ensure singlino domination. Figure \ref{fig:relic}(a) illustrates
how the massless or very light neutralinos of $M_{\chi^0_1} < 1$ GeV
can appear without violating direct constraints.

The $M_{\chi^0_1}$ upper bound is $83$ GeV at $\tan\beta = 1$, which
is considerably lower than those of other models\footnote{Including
values with $M_i < 0$, the $M_{\chi^0_1}$ upper bound can increase,
but only up to $100$ GeV.}; the bound is due to the upper limit
imposed on $h_s$, and the bound significantly increases without this
constraint. Figure \ref{fig:s}(a) shows that $M_{\chi^0_1}$
decreases with $s$, as we expect from Eq. (\ref{detm}). As seen in
Figure \ref{fig:s}(b), the singlino component in $\chi^0_1$
increases with increasing $s$ and becomes dominant for sufficiently
large $s$. While $M_{\chi^0_1}$ increases with decreasing $s$, $s$
cannot be too small because of the $h_s \le 0.75$ requirement. The
singlino composition at the maximum $M_{\chi^0_1}$ is still dominant
with $|N_{15}|^2$ of $0.7$. When there are other sizable components
in $\chi^0_1$, the $\Gamma_{Z \to \chi_1^0 \chi_1^0}$ constraint is
not easy to escape unless $\tan\beta \simeq 1$ where the
$Z$-$\chi_1^0$-$\chi_1^0$ coupling (Eq. (\ref{eqn:Zwidth}))
vanishes. At $\tan\beta = 1$ (or $v_1 = v_2$), the contributions of
the two doublet Higgsinos to $\chi^0_1$ are identical (up to sign)
as Eq. (\ref{eqn:massmatrixNMSSM}) suggests.

Because of the small $\chi^0_1$ mass in this model, most
annihilation channels for the MSSM $\chi^0_1$ relic density
calculation become irrelevant, and the $Z$ pole is the dominant
channel \cite{neutralino_nMSSM, relic_U1}. Figure \ref{fig:relic}(b)
shows the neutralino relic density in this model through the $Z$
pole annihilation. The direct constraints of Section \ref{direct}
are applied. To reproduce the acceptable relic density with only the
$Z$ pole annihilation contribution, the lower $M_{\chi^0_1}$
bound\footnote{The bound can be lowered a bit for a weaker
$\Gamma_{Z \to \chi_1^0 \chi_1^0}$ constraint.} is $M_{\chi^0_1}
\gsim 30$ GeV, while the upper bound remains the same as discussed
above. This is the most severe lower bound on $M_{\chi^0_1}$ in the
model. However, the Higgs masses are not bounded by the LEP2 data
because of possible mixing between Higgs doublets and singlets, and
very light Higgses may provide sufficiently large annihilation so
that even lighter neutralinos are allowed by the relic density
constraint. For an explicit calculation of such a light neutralino
relic density through a light pseudoscalar state, see Ref.
\cite{lightneutralino}. We also refer to Ref.
\cite{lightpseudoNMSSM} for interesting physics associated with a
light pseudoscalar Higgs boson.

As can be seen in Figure \ref{fig:tanb}(a), the lightest neutralinos
that are massive enough for the annihilation through the $Z$ pole
($M_{\chi^0_1} \gsim 30$ GeV) are allowed only for small
$\tan\beta$. This raises concern since the Supersymmetric
contribution to $(g-2)_\mu$ is proportional to $\tan\beta$, and a
large value of $\tan\beta$ is favored to explain the considerable
$2.4\sigma$ deviation of the experimental value from the SM
prediction. Nonetheless, a common solution was found to exist that
can explain both the acceptable relic density through the $Z$ pole
and the deviation of $(g-2)_\mu$ in this model, as illustrated in
Figure \ref{fig:histogram}(a) \cite{g-2_U1}.

\subsection{In the UMSSM}
In the UMSSM, the upper bound on the lightest neutralino mass is
typically given by the maximal value of $M_1$ (with $s\gg M_1$),
where $\chi^0_1$ is essentially the MSSM-like neutralino with almost
no singlino composition. The $\chi^0_1$ mass bound does not increase
above $420$ GeV when the $M_{1,2}$ ranges are extended, unlike the
MSSM case (Figure \ref{fig:M2}(b)). This is because of our
restriction $s \le$ 2000 GeV: for the extended $M_{1,2}$ range one
is in a regime similar to Eq. (\ref{eqn:massmatrix56b}), with
$M_{\chi^0_1}$ given by the smaller of the eigenvalues in Eq.
(\ref{eqn:mmeigns}).
As the solid curves of Figure \ref{fig:s} illustrate, both the mass
and the singlino composition of the lightest neutralino increase
with $s$ until $s$ reaches a certain value where that neutralino is
no longer the lightest and the singlino composition of $\chi^0_1$
plunges to zero.

A vanishing lower bound on $M_{\chi^0_1}$ occurs at $\tan\beta = 1$
and $s = 365$ GeV with a singlino composition of $|N_{15}|^2 \simeq
0.6$ and a $Z'$-ino composition of $|N_{16}|^2 \simeq 0.2$. Though
the singlino component does not saturate $\chi^0_1$ here, the
next-to-dominant component, $Z'$-ino, also does not contribute to
the $Z$ decay width, and such a light neutralino survives the
$\Gamma_{Z \to \chi_1^0 \chi_1^0}$ constraint.

The Tevatron lower bound on $M_{Z'}$ is too large for the $Z'$ to
make a significant contribution to $(g-2)_\mu$ through a $Z'$-loop,
though the $\mu$-$\chi^0_i$-$\tilde \mu_j$ contribution is modified
by the $Z'$-ino component in $\chi^0_i$ \cite{g-2_U1}. Figure
\ref{fig:M2}(a) shows the parameter space constrained by
$(g-2)_\mu$ for two values of the smuon mass.

A value of $s \sim \cal{O}$(EW) can be rather problematic, as
discussed in Section \ref{sec:Z'mass}, unless either there are
additional contributions to $M_{Z'}$ (S-model) or the leptonic
coupling is small (leptophobic $Z'$ model). If we take a TeV scale
lower bound on $s$ (for example, in the $s \gsim 1.5$ TeV range in
Figure \ref{fig:s}) to satisfy the Tevatron $M_{Z'}$ bound of $500
\sim 800$ GeV, the singlino and $Z'$-ino components are negligible
and the lightest neutralino mass becomes similar to that of the
MSSM.

\section{Summary and Conclusion}
\label{conclusion}
\begin{figure}[t]
\begin{minipage}[b]{1\textwidth}
\begin{minipage}[b]{.49\textwidth}
\centering\leavevmode \epsfxsize=3in \epsfbox{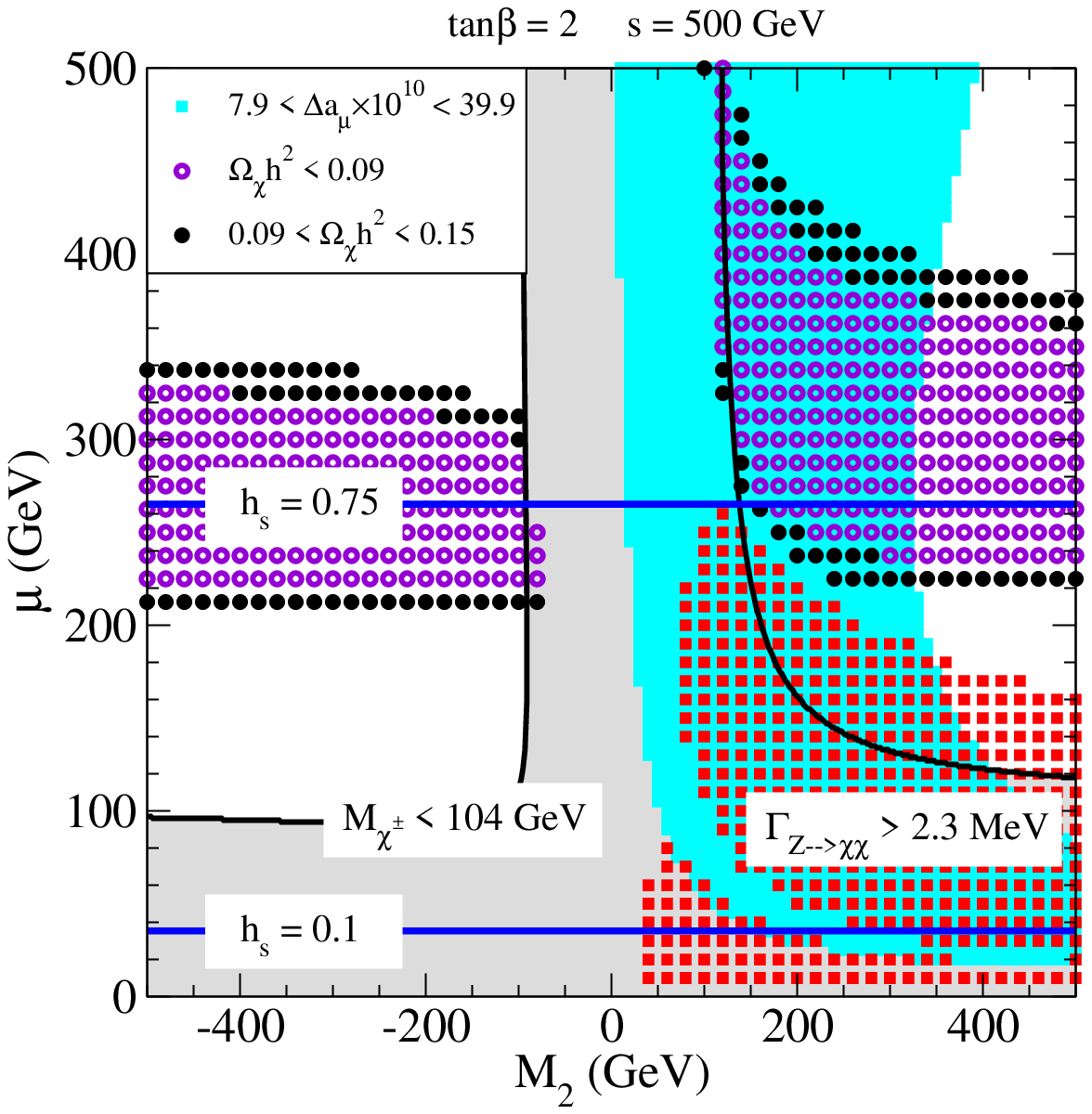}
\end{minipage}
\hfill
\begin{minipage}[b]{.49\textwidth}
\centering\leavevmode \epsfxsize=3in \epsfbox{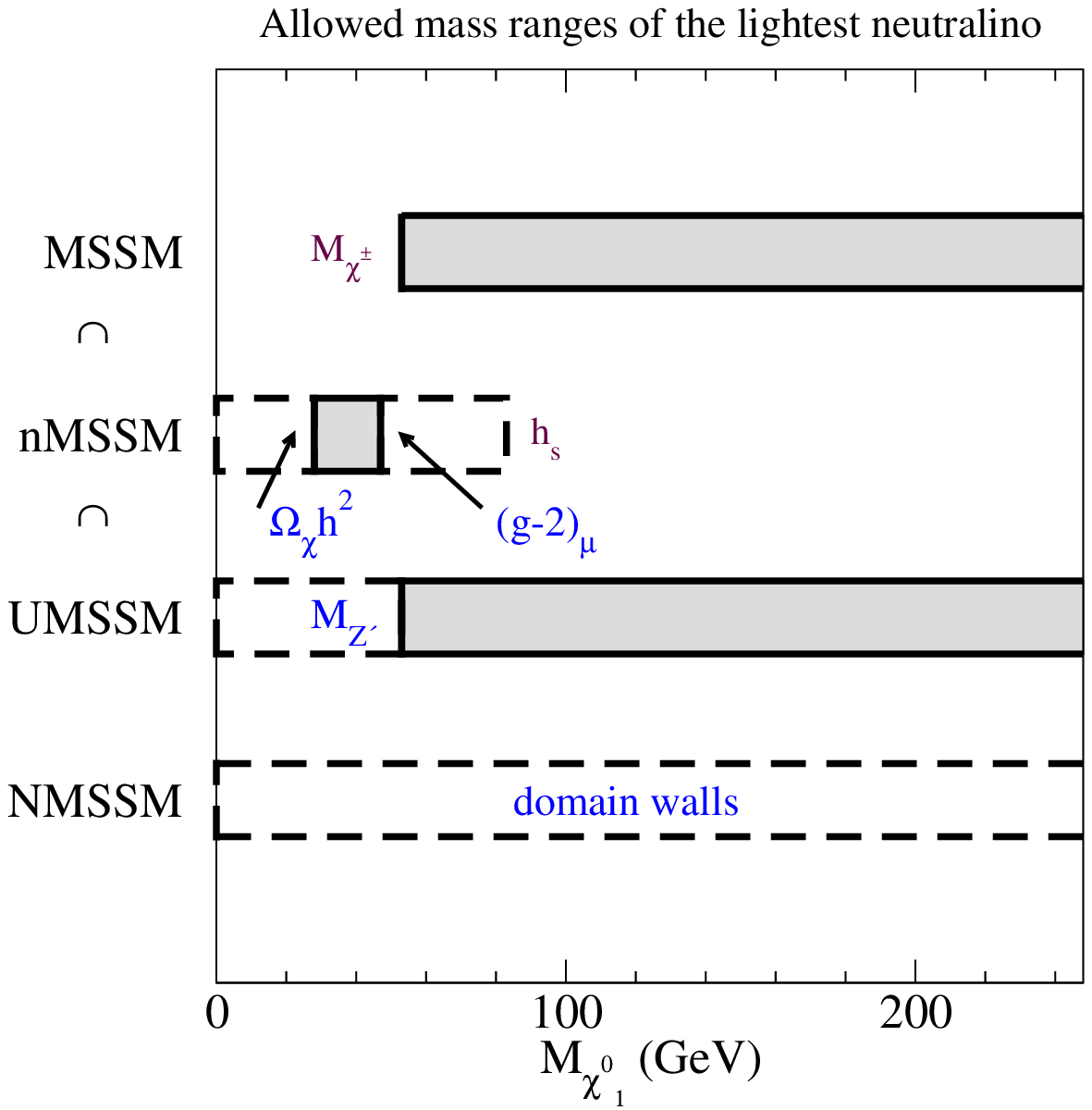}
\end{minipage}
(a)\hspace{0.48\textwidth}(b)
\end{minipage}
\caption{ (a) An illustration of a common solution of the
$\Omega_{\chi^0_1} h^2$ (dark dots, $3 \sigma$ of WMAP+SDSS) and the
$(g-2)_\mu$ (cyan region, $95\%$ C.L. of E821) in the $M_2$-$\mu$
plane along with other constraints in the nMSSM or the S-model. The
$Z$-pole annihilation and $m_L = m_E = 100$ GeV are used. The gap in
the $\Gamma_Z$ exclusion region is where $|N_{13}|^2 \approx
|N_{14}|^2$. (b) The allowed mass range of $\chi^0_1$ after applying
direct constraints (model parameters, gaugino mass unification
$M_{1'} = M_1 \simeq 0.5 M_2$, $M_{\chi^\pm_1}$, $\Delta \Gamma_Z$)
and indirect constraints ($(g-2)_\mu$, $\Omega_{\chi^0_1} h^2$,
$M_{Z'}$, domain wall). The $M_{\chi^0_1}$ bounds in the nMSSM are
intended to be illustrative and are not necessarily quantitatively
precise.} \label{fig:histogram}
\end{figure}

Although Supersymmetry at the TeV scale is well-motivated, the MSSM
is just one of the possible realizations. In fact, the theoretical
$\mu$-problem suggests that the MSSM is incomplete. The solution to
the $\mu$-problem suggests that an appropriate direction to extend
the MSSM is to have an extra Higgs singlet whose VEV gives the
effective $\mu$-term of the EW scale. Extensions of the MSSM have
extra neutralinos, and the composition of the lightest neutralino
involves extra components beyond those of the MSSM. Because of this,
both the mass and couplings of the lightest neutralino are modified
from the MSSM. The lightest neutralino $(\chi^0_1)$ is interesting
both in particle physics (as the LSP) and cosmology (as the CDM),
and it is therefore important to study and compare properties of the
$\chi^0_1$ in extended MSSM models.

We explored $\chi^0_1$ properties in various extended MSSM models.
We examined constraints from the experimental bound on the $Z \to
\chi^0_1 \chi^0_1$ contribution to $\Gamma_Z$, the lower bound on
$M_{\chi_1^\pm}$, and constraints from perturbativity and
naturalness of the  $S H_1 H_2$ and $S^3$ coupling strengths. We
also considered constraints from the relic density of the CDM
($\Omega_{\chi^0_1} h^2$), the experimental deviation of $(g-2)$ of
the muon from the SM expectation, and the Tevatron lower bound on
$M_{Z'}$. Distinguishing properties of the lightest neutralino arise
in extended MSSM models, such as distinct $\chi^0_1$ mass ranges
(Eq. (\ref{eqn:MSSMrange}) - (\ref{eqn:UMSSMrange})), frequent
singlino dominance (Figure \ref{fig:tanb}(b)), importance of the $Z$
pole annihilation channel in the relic density calculation (Figure
\ref{fig:relic}(b)) and different $\tan\beta$ dependences of the
upper and lower mass bounds (Figure \ref{fig:tanb}(a)). Some of the
extended MSSM models can be considered as limits of the other
models. For example, as far as the neutralino sector is concerned,
we can consider that MSSM $\subset$ nMSSM $\subset$ UMSSM $\subset$
S-model.

Approximate lightest neutralino mass ranges in the models considered
are illustrated in Figure \ref{fig:histogram}(b). The dashed regions
are disfavored by the indirect constraints. After the $M_{Z'}$ lower
bound is imposed, the UMSSM bound becomes similar to the MSSM. In
the case of the nMSSM there is a tension between the $(g-2)_\mu$
constraint which favors small $M_{\chi^0_1}$ (or large $\tan\beta$)
and the relic density constraint which favors large $M_{\chi^0_1}$
(or small $\tan\beta$).

The properties of the CDM particle, even if it is the lightest
neutralino, may be quite different from the MSSM prediction. For
instance, it could be extremely light and/or dominated by the
singlino, which does not directly couple to SM particles except
Higgs doublets. Similar distinctions of models may occur in the
Higgs sector.

Even if a low energy Supersymmetry is correct, its realization may
depend on the model. The measurement of the mass of the lightest
neutralino and the determination of its couplings will be
particularly useful in testing the MSSM and its extensions at
colliders.

\section*{Acknowledgments}
HL thanks K. Matchev and the High Energy Theory Group at University
of Florida for hospitality and discussions during a visit. HL also
thanks D. Morrisey for a useful discussion. This research was
supported in part by the U.S. Department of Energy under Grants
No.~DE-FG02-95ER40896 and No.~DOE-EY-76-02-3071, and in part by the
Wisconsin Alumni Research Foundation.


\end{document}